\documentclass[aps, prd, 12pt, eqsecnum, tightenlines, notitlepage, nofootinbib, floatfix]{revtex4-1}

\PassOptionsToPackage{unicode}{hyperref}

\usepackage{hyperref, graphicx, multirow, slashed, xcolor, amsmath, breakurl}

\allowdisplaybreaks

\definecolor{red}{rgb}{1,0,0}

\begin{document}

\title{Static Spherically Symmetric Kerr-Schild Metrics and Implications for the Classical Double Copy}

\author{Alexander K. Ridgway}
\author{Mark B.\ Wise}
\affiliation{Walter Burke Institute for Theoretical Physics,
California Institute of Technology, Pasadena, CA 91125}

\begin{abstract}

We discuss the physical interpretation of  stress-energy tensors that source static spherically symmetric Kerr-Schild metrics.  We find that the sources of such metrics with no curvature singularities or horizons do not simultaneously satisfy the weak and strong energy conditions.  Sensible stress-energy tensors usually satisfy both of them.  Under most circumstances these sources are not perfect fluids and contain shear stresses.  We show that for these systems the classical double copy associates the electric charge density to the Komar energy density.  In addition, we demonstrate that the stress-energy tensors are determined by the electric charge density and their conservation equations.

\end{abstract}

\maketitle

\section{Introduction}

In the last 20 years, amplitude based methods have become important in the study of gauge theories and gravity.  Apart from the ease they provide us with carrying out explicit perturbative calculations in gauge theories, amplitudes have revealed hidden structures in them that are not transparent in the local Lagrangian formulation.  A prime example of this is the BCJ duality and perturbative double copy introduced in \cite{Bern:2008qj} which reveals a deep connection between perturbative gauge theory and gravity amplitudes.  This double copy states that one can deduce the $n$-point graviton amplitude straight from the $n$-point Yang-Mills one if the Yang-Mills amplitude is written in BCJ form (see ref. \cite{Bern:2008qj}, \cite{Bern:2010yg}, \cite{Bern:2010ue}, and \cite{Monteiro:2011pc}).  This duality is proven at tree level and expected to hold at all orders in perturbation theory.

While the perturbative double copy is well understood operationally, there are still many unanswered questions on the subject.  For example, is the double copy relationship between gauge and gravity theories only perturbative or does it manifest itself in other aspects of the theories?  Investigating along these lines, the authors of \cite{Monteiro:2014cda} found that the classical solutions of General Relativity and Electromagnetism exhibit similar double copy structure.  Specifically, they found that time-independent Kerr-Schild metrics can be interpreted as the ``classical double copies" of time-independent E\&M vector potentials.  

We now give a couple of illustrative examples (see \cite{Monteiro:2014cda}).  The first example is that the Kerr-Schild form of the Schwarzschild metric, which is sourced by a point mass \cite{Balasin:1993kf}, is the double copy of the electrostatic potential for a point charge.  The second example is that the Kerr-Schild form of the Kerr metric, which is sourced by a rotating disc of mass and the pressures needed to stabilize it \cite{Israel:1970kp}, is the double copy of the vector potential of a rotating disc of charge.  

These two examples underscore a couple important aspects about the classical double copy of static E\&M vector potentials.  The classical double copy relates the energy-momentum distribution sourcing the metric to the  charges and currents sourcing the  E\&M vector potential.  Furthermore the sources for the metric generally include pressures that are not present in the E\&M source.  These additional terms provide forces that stabilize the source distribution and make the metric time-independent.  

In this note we analyze static spherically symmetric Kerr-Schild metrics with no curvature singularities and no horizons\footnote{Indeed, there are many simple examples of metrics that have singularities and/or horizons, such as the Schwarzschild and de-Sitter metrics.  For simplicity, however, we do not include them in our analysis.}.  Our goal is to understand the stress-energy tensors that source these metrics and determine the relationship between these stress-energy tensors and the sources of the corresponding E\&M problem.  As one might expect, the classical double copy associates static spherically symmetric Kerr-Schild metrics with spherically symmetric electrostatic potentials.  It also naturally relates the charge density of the E\&M problem to the Komar energy density of the gravity one.  

We show that the stress-energy tensors sourcing these double copies are usually not perfect fluids and contain shear stresses.  Furthermore, we conclude the GR strong and weak energy conditions cannot simultaneously be satisfied by these stress-energy tensors.  We also show that the stress-energy tensors sourcing spherically symmetric double copies are composed of a piece fixed by the sources of the E\&M problem and another piece determined by conservation of the stress-energy tensor.  This is consistent with the examples presented in \cite{Monteiro:2014cda}.

The paper is organized as follows.  In section 2 we review Kerr-Schild metrics and the classical double copy presented in \cite{Monteiro:2014cda}.  In section 3 we deduce the stress-energy tensor of a static spherically symmetric Kerr-Schild metric from Einstein's equations and interpret its components.  In section 4 we re-derive the stress-energy tensor found in section 3 using the classical double copy and conservation of the stress-energy tensor.  In section 5 we test these stress-energy tensors against the GR energy conditions and examine their physical properties.  We end with concluding remarks in section 6.

\section{The Classical Double Copy}
In this section we summarize the double copy for classical solutions (which we call the classical double copy) first presented in \cite{Monteiro:2014cda}.  We begin by defining a Kerr-Schild metric as a metric whose components can be written in some coordinate system \cite{Debney:1969zz} (see also \cite{Stephani}), which we call Kerr-Schild coordinates, as
\begin{align}
\label{metric}
g_{\mu\nu} = \eta_{\mu\nu} + \phi k_{\mu}k_{\nu}
\end{align}
where $\eta_{\mu\nu} = \textrm{diag}\left(-1,1,1,1\right)$ is the Minkowski metric and $k^{\mu}$ is a geodesic null vector satisfying
\begin{align}
\label{Kerr vec}
k^{2} &= 0\cr
 k^{\mu}\partial_{\mu}k_{\nu} &= 0.
\end{align}
The fact that $k^{\mu}$ squares to zero implies the components of the inverse metric can be expressed
\begin{align}
\label{inverse metric}
g^{\mu\nu} = \eta^{\mu\nu} - \phi k^{\mu}k^{\nu}.
\end{align}
By equations (\ref{metric}) and (\ref{inverse metric}), it is equivalent to raise and lower $k^{\mu}$'s index with either the Minkowski or the full metric.  Let us adopt the convention to also raise and lower a partial derivative's index with the Minkowski metric.  With this convention the Ricci tensor with one index raised and one lowered can be written
\begin{align}
\label{Ricci tensor}
R^{\mu}{}_{\nu} &= \frac{1}{2}\left(\partial^{\mu}\partial_{\alpha}\left(\phi k^{\alpha}k_{\nu}\right) + \partial_{\nu}\partial_{\alpha}\left(\phi k^{\alpha}k^{\mu}\right) - \partial^{2}\left(\phi k^{\mu}k_{\nu}\right)\right).
\end{align} 
Let's now specialize to the case of time-independent metrics: $\partial_{0}\phi = \partial_{0}k_{\nu} = 0$.  We also set $k^{0} = 1$ by absorbing any functional dependence $k^{0}$ might have into $\phi$.  Then since $k^{\mu}$ squares to zero we have
\begin{align}
\label{general Kerr vector}
 k^{\mu} = \left(1, \hat{k}\right)
 \end{align}
where $\hat{k}$ is a unit 3-vector.  By plugging (\ref{general Kerr vector}) into (\ref{Ricci tensor}) we can express the Ricci tensor's components as
\begin{align}
\label{Ricci comps}
R^{\mu}{}_{0} &= -\frac{1}{2}\partial_{\nu}\left(\partial^{\mu}\left(\phi k^{\nu}\right) - \partial^{\nu}\left(\phi k^{\mu}\right)\right)\cr
R^{i}{}_{j} &= \frac{1}{2}\partial_{l}\left(\partial^{i}\left(\phi k^{l}k_{j}\right) + \partial_{j}\left(\phi k^{l}k^{i}\right) - \partial^{l}\left(\phi k^{i}k_{j}\right)\right).
\end{align}
In the context of the classical double copy the Kerr-Schild metric (\ref{metric}) is referred to as the ``double copy" since it involves two factors of the vector $k^{\mu}$.  By removing one copy of $k^{\mu}$ from (\ref{metric}) we can construct the ``single copy" and its associated field strength
\begin{align}
A^{\mu} &= \phi k^{\mu}\cr
F^{\mu\nu} &= \partial^{\mu}A^{\nu} - \partial^{\nu}A^{\mu}.
\end{align}
Using these definitions we write the first equation in (\ref{Ricci comps}) as
\begin{align}
\label{Maxwell}
R^{\mu}{}_{0} = -\frac{1}{2}\partial_{\nu}F^{\mu\nu} = -2\pi j^{\mu}.
\end{align}
We immediately recognize (\ref{Maxwell}) as the sourced Maxwell's equations.  This means we can interpret the single copy $A^{\mu}$ as the E\&M vector potential sourced by the the current 
\begin{align}
j^{\mu} = -\frac{1}{2\pi}R^{\mu}{}_{0}.
\end{align}
It is difficult to see such a clear relationship between the Ricci tensor's spatial components and $j^{\mu}$ in the second equation of (\ref{Ricci comps}).  However, by using (\ref{Ricci tensor}) and the properties of $k^{\mu}$ it can be shown
\begin{align}
\label{R identity}
R^{\mu}{}_{\nu}k^{\nu}k_{\mu} = 0.
\end{align}
By expanding out the contractions in (\ref{R identity}) we can write
\begin{align}
\label{spatial connection}
R^{\hat{k}}{}_{\hat{k}} = -2 \pi \left( \rho - 2{\vec j }\cdot\hat{k} \right)
\end{align}
where we've defined $\rho = j^{0}$ and $R^{\hat{k}}{}_{\hat{k}} = R^{i}{}_{j}\hat{k}^{j}\hat{k}_{i}$.  The spatial component $R^{\hat{k}}{}_{\hat{k}}$ is fixed by $\rho$ and the component of the spatial current parallel to $\hat{k}$.  	

Equations (\ref{Maxwell}) and (\ref{spatial connection}) embody the classical double copy relationship between $A^{\mu}$ and $g_{\mu\nu}$.  In general the other Ricci components will depend on terms parameterized by $\phi$, $\hat{k}$ and their derivatives.  Typically these terms will not have simple relationships to the components of $j^{\mu}$ like the $R^{\mu}{}_{0}$ and $R^{\hat{k}}{}_{\hat{k}}$ components did in (\ref{Maxwell}) and (\ref{spatial connection}).

\section{Static Spherically Symmetric Kerr-Schild Metrics}
We now restrict our attention to static spherically symmetric Kerr-Schild metrics.  We will interpret the components of the stress-energy tensors sourcing these metrics and relate them to the sources of the corresponding single copies.  The most general spherically symmetric Kerr-Schild solution that can be written is
\begin{align}
\label{spherical metric}
g_{\mu\nu} = \eta_{\mu\nu} + \phi (r)k_{\mu}k_{\nu}
\end{align}
where $k^{\mu} = \left(1, \hat{r}\right)$.  It is easily seen that this is the only spherically symmetric $k^{\mu}$ that satisfies constraints (\ref{Kerr vec}) and $k^{0} = 1$.  The single copy of this metric is
\begin{align}
A^{\mu} = \phi(r)\left(1, \hat{r}\right).
\end{align}
One can perform the gauge transformation $A^{\mu} \rightarrow A^{\mu} + \partial^{\mu}\lambda$ where 
\begin{align}
\lambda = -\int^{r}\phi(r') dr'
\end{align}
to write $A^{\mu} = \phi(r)\left(1, 0,0,0\right)$.  This means the single copy of a spherically symmetric Kerr-Schild metric is an electrostatic potential sourced by a spherically symmetric distribution of charge
\begin{align}
j^{\mu} = \rho(r)\left(1,0,0,0\right)
\end{align}
where as before, $\rho(r) = -{\nabla^{2}\phi}/{4\pi}$.  The Ricci tensor components for (\ref{spherical metric}) are
\begin{align}
\label{Spherical Ricci}
R^{0}{}_{0} &= \frac{\nabla^{2}\phi}{2},\ R^{0}{}_{i} = R^{i}{}_{0} = 0\cr
R^{i}{}_{j} &= \frac{1}{2}\left[\nabla^{2}\phi\left(\frac{x^{i}x_{j}}{r^{2}}\right) + 2\left(\frac{\phi'}{r} + \frac{\phi}{r^{2}}\right)\left(\delta^{i}{}_{j} - \frac{x^{i}x_{j}}{r^{2}}\right)\right].
\end{align}
As expected, $R^{0}{}_{0}$ and $R^{\hat{r}}{}_{\hat{r}}$ are parameterized by the E\&M charge density $\rho$ and since $j^{i} = 0$, $R^{i}{}_{0} = 0$.  On the other hand, the components of $R^{i}{}_{j}$ perpendicular to $\hat{r}\hat{r}$ depend on the combination
\begin{equation}
\label{mass parameter eq}
{\tilde \rho} ={ {\phi'} \over {r}} + {{\phi}\over {r^{2}}} 
\end{equation}
rather than $j^{0}=\rho$.  Before we talk more about ${\tilde{ \rho}}$ lets interpret $\rho$ on both sides of the classical double copy.  On the E\&M side $\rho$ is the charge density sourcing the vector potential $A^{\mu}$.  It turns out that it can be interpreted on the gravity side as the Komar energy density \cite{Komar} (and see for example \cite{Sean's book}). 

It is useful to change to a coordinate system in which $g_{\mu\nu}$ is diagonal.  If we use unprimed coordinates to denote the old coordinate system and primed ones to denote the new ones then the coordinate transformation that diagonalizes the metric is
\begin{align}
\label{diag trans}
dt &= dt' - \frac{\phi}{1 - \phi}dr'\cr
dr &= dr'.
\end{align}
The metric components in the new coordinate system, which we call Schwarzschild coordinates, are
\begin{align}
\label{Schwarzschild metric}
g_{\mu\nu} = \textrm{diag}\left(-\left(1 - \phi\right),\frac{1}{1 - \phi}, r^{2}, r^{2}{\rm sin}^{2}\theta\right).
\end{align}
It is important to note that (\ref{Schwarzschild metric}) is not the most general spherically symmetric metric.  The fact that $g_{rr}$ is minus the inverse of $g_{tt}$ is a consequence of the metric being Kerr-Schild.  We focus on metrics where $\phi(r) \textless 1$ to ensure there are no horizons.  We will also assume there are no curvature singularities.

Let's choose units such that $c = \hbar = 8\pi G = 1$.\footnote{If you keep the units for G, then the electromagnetic charge density is $\sqrt{8\pi G}\rho_{K}$.}  We now calculate the Komar energy in the Schwarzschild frame.  Since we have a timelike Killing vector $K^{\mu} = \left(1,0,0,0\right)$, we can express the Komar energy contained within a 3-dimensional spatial volume $\Sigma$ as
\begin{align}
\label{Komar}
E = 2\int_{\Sigma}d^{3}x\sqrt{\gamma}n_{\mu}K^{\nu}R^{\mu}{}_{\nu}
\end{align}
where $\gamma ={r^{4}{\rm sin}^{2}\theta}/(1 - \phi)$ is the determinant of the induced metric on the volume $\Sigma$ and $n^{\mu} =\left(1/\sqrt{1 - \phi}, {\vec 0} \right)$ is the unit normal to the surface.  The $R^{0}{}_{0}$ component doesn't change under the coordinate transformation (\ref{diag trans}) so we can then plug the first equation of (\ref{Spherical Ricci}) into (\ref{Komar}) to find
\begin{align}
E = -\int_{\Sigma}r^{2}drd\Omega \nabla^{2}\phi.
\end{align}
The Komar energy density is then
\begin{align}
\rho_{K} = - \nabla^2 \phi=4\pi \rho
\end{align}
The classical double copy naturally associates the charge density $\rho$ to the Komar energy density $\rho_{K}$.   We restrict ourselves to potentials such that $\rho_{K}$ and $\tilde{\rho}$ are both finite.  This prevents scalars such as $R^{\mu}_{~\nu}R_{\mu}^{~\nu}= \rho_K^2/2+2{\tilde{ \rho}}^2$ from exhibiting curvature singularities.  Hence we impose
\begin{align}
\label{finite constraints}
\phi(0) = \phi'(0) = 0.
\end{align}
Let us now interpret ${\tilde \rho}$ by matching the stress-energy tensor sourcing the double copy to a general fluid solution.  Using Einstein's equations, we solve for the stress-energy tensor in terms of the Ricci tensor.  In our units
\begin{align}
\label{Einstein}
R^{\mu}{}_{\nu} - \frac{1}{2}\delta^{\mu}{}_{\nu}R = T^{\mu}{}_{\nu}.
\end{align}
By plugging (\ref{Spherical Ricci}) into (\ref{Einstein}) we can find the components of the stress-energy tensor with one index raised and one lowered in Kerr-Schild coordinates.  We then transform to Schwarzschild coordinates using (\ref{diag trans}) and express the stress-energy tensor with two lowered indices using (\ref{Schwarzschild metric}).  The result is
\begin{align}
\label{diagonal stress energy}
T_{\mu\nu} &= \textrm{diag}\left(\left(1 - \phi\right)\left(\frac{\phi'}{r} + \frac{\phi}{r^{2}}\right), -\frac{1}{1 - \phi}\left(\frac{\phi'}{r} + \frac{\phi}{r^{2}}\right), -\frac{r^{2}\nabla^{2}\phi}{2}, -\frac{r^{2}{\rm sin}^{2}\theta\nabla^{2}\phi}{2}\right)\cr
&= \left(\left(1 - \phi\right){ \tilde{\rho}}, -\frac{1}{1 - \phi}{ \tilde{\rho}}, \frac{r^{2}}{2}\rho_{K}, \frac{r^{2}{\rm sin}^{2}\theta}{2}\rho_{K}\right).
\end{align}
We  interpret these components by matching them to the components of a general fluid solution. This stress-energy tensor is characterized by matter energy density, pressures, shear stress and heat flux,
\begin{align}
\label{fluid solution}
T_{\mu\nu} = \left(\rho_{m} + p_m\right)u_{\mu}u_{\nu} + p_{m}g_{\mu\nu} + \left(u_{\mu}q_{\nu} + u_{\nu}q_{\mu}\right) + \pi_{\mu\nu}
\end{align}
where $\rho_{m}$  is the matter energy density, $p_m$ is the pressure stabilizing the matter, $u^{\mu}$ is the four-velocity of the fluid, $q^{\mu}$ is the heat flux and $\pi_{\mu\nu}$ is the shear tensor which satisfies $\pi_{\mu\nu}u^{\mu} = 0$ and $g^{\mu \nu}\pi_{\mu\nu} = 0$ .  Matching (\ref{diagonal stress energy}) to (\ref{fluid solution}) gives
\begin{align}
\label{heatless fluid}
T_{\mu\nu} = \left(\rho_{m} + p_m\right)u_{\mu}u_{\nu} + p_{m} g_{\mu\nu} + \pi_{\mu\nu}
\end{align}
where
\begin{align}
\label{fluid comps}
q^{\mu} &= 0,\ \rho_{m}={\tilde \rho} ,\ p_m = \frac{1}{3}\left(\rho_{K} - \rho_{m}\right)\cr
\pi_{\mu\nu} &= \pi_{ij} = -\frac{3}{2}\left(\rho_{m} + p_m\right)\left(\frac{3 - \phi}{3\left(1 - \phi\right)}\frac{x_{i}x_{j}}{r^{2}} - \frac{1}{3}\delta_{ij}\right).
\end{align}
The first thing to note is that the classical double copy does not always yield perfect fluids.  In fact, since
\begin{equation}
\rho_m+p_m= {1 \over 3}\left( {2\phi \over r^2}- \phi'' \right)
\end{equation}
only for $\phi \propto r^{2}$ or $\phi \propto 1/r$ does the shear term vanish and a perfect fluid source the double copy.  The $1/r$ case is the potential of a point charge, which gets double copied to the Schwarzschild metric, while the $r^2$ case corresponds to a constant charge density and double copies to de-Sitter space-time and anti de-Sitter space-time.   Shear stresses will be needed to stabilize all other spherically symmetric Kerr-Schild solutions.

Another thing to note is that $\rho_{m} = \tilde{\rho}$ is the energy density of the fluid and, as can be seen from (\ref{mass parameter eq}), is not directly related to the E\&M source $j^{\mu}$.  The pressure is proportional to the difference between the Komar energy density and $\rho_{m}$.  This implies the Komar energy gets contributions from the energy due to the stabilizing isotropic pressure $p_{m}$ and the energy density $\rho_{m}$.  The double copy then maps the electric charge density to a linear combination of $\rho_{m}$ and internal pressures.

\section{The Stress Tensor from Conservation and E\&M sources}
Earlier we mentioned the stress-energy tensor has a piece determined by the sources of the single copy and another piece that stabilizes the distribution.  We now show this explicitly using conservation of the stress-energy tensor.  The general form of the Ricci tensor for a spherically symmetric source is
\begin{align}
R^{0}{}_{0} &= A(r),\ R^{i}{}_{0} = B(r)\frac{x^{i}}{r}\cr
R^{i}{}_{j} &= C(r)\frac{x^{i}x_{j}}{r^{2}} + D(r)\left(\delta^{i}{}_{j} - \frac{x^{i}x_{j}}{r^{2}}\right).
\end{align}
Let's now specialize to Kerr-Schild metrics.  By equations (\ref{Maxwell}) and (\ref{spatial connection}), the $R^{\mu}{}_{0}$ and $R^{\hat{r}}{}_{\hat{r}}$ components are determined by the single copy's sources.  So $R^{\mu}{}_{\nu}$ can be written as
\begin{align}
\label{Other ricci stuff}
R^{0}{}_{0} &= \frac{\nabla^{2}\phi}{2},\ R^{i}{}_{0} = 0\cr
R^{i}{}_{j} &= \frac{\nabla^{2}\phi}{2}\frac{x^{i}x_{j}}{r^{2}} + D(r)\left(\delta^{i}{}_{j} - \frac{x^{i}x_{j}}{r^{2}}\right).
\end{align}
We can fix $D(r)$ using conservation of the stress-energy tensor.  Plugging (\ref{Other ricci stuff}) into Einstein's equations, we find
\begin{equation}
\label{Stress energy with D}
T^{0}{}_{0} = -D(r)\ {\rm and}\ T^{i}{}_{j} = -\frac{\nabla^{2}\phi}{2}\left(\delta^{i}{}_{j} - \frac{x^{i}x_{j}}{r^{2}}\right) - D(r)\frac{x^{i}x_{j}}{r^{2}}
\end{equation}
The components (\ref{Stress energy with D}) turn out to be the same in both Kerr-Schild and Schwarzschild coordinates.  Let's now impose conservation of stress-energy
\begin{align}
\label{conservation of stress E}
\nabla_{\mu}T^{\mu}{}_{\nu} = 0.
\end{align}
In stars, for example, it is (\ref{conservation of stress E}) that gives rise to the equation for hydrostatic equilibrium.  Since we're working in Kerr-Schild coordinates this equation simplifies and becomes
\begin{align}
\partial_{\mu}T^{\mu}{}_{\nu}	- \frac{1}{2}\partial_{\nu}\left(\phi k^{\alpha}k_{\mu}\right)T^{\mu}{}_{\alpha} = 0.
\end{align}
By time independence the $\nu = 0$ component is satsified trivially.  The $\nu = k$ component reduces to the constraint
\begin{align}
\nabla^{2}\phi - rD'(r) - 2D(r) = 0.
\end{align}
The solution to this equation is
\begin{align}
D(r) = \left(\frac{\phi'}{r} + \frac{\phi}{r^{2}}\right) + \frac{d}{r^{2}}
\end{align}
where $d$ is a constant.  To avoid curvature singularities at the origin we set $d = 0$.  We have reproduced the Ricci and stress-energy components obtained in the previous section.  This calculation shows that the piece of the stress-energy tensor not fixed by the single copy is responsible for stabilizing the gravitational system.  This is consistent with the Schwarzschild and Kerr metrics presented in \cite{Monteiro:2014cda} which exhibit stress-energy tensors composed of an E\&M piece and a stabilization one.

\section{The Energy Conditions}
In this section we review some of the GR energy conditions.  Most sensible forms of matter should satisfy these energy conditions\footnote{An exception is vacuum energy which can have either sign for $\rho_{m}$ and has $\rho_{m} + p_{m} = 0$.}.  Let us start with the null energy condition, which places the fewest constraints on the source.  The null energy condition demands
\begin{align}
T_{\mu\nu}l^{\mu}l^{\nu} \geq 0
\end{align}
where $l^{\mu}$ is any null vector.  For (\ref{heatless fluid}) this implies
\begin{align}
\label{null E}
\left(\rho_{m} + p_{m}\right)\left(u_{\mu}l^{\mu}\right)^{2} + \pi_{\mu\nu}l^{\mu}l^{\nu}\geq 0.
\end{align}
We interpret (\ref{null E}) below when we specialize to the sources of spherically symmetric Kerr-Schild metrics. 

Next we consider the so-called weak energy condition which demands that the stress-energy tensor satisfies
\begin{align}
\label{weak energy}
T_{\mu\nu}t^{\mu}t^{\nu}\geq 0
\end{align}
where $t^{\mu}$ is any timelike vector.  This condition ensures that observers moving along timelike paths observe positive energy densities.  To derive the conditions (\ref{weak energy}) places on the components of (\ref{heatless fluid}) we first note that any timelike vector $t^{\mu}$ can be expressed as a linear combination of the fluid's four velocity $u^{\mu}$ and a null vector $l^{\mu}$:
\begin{align}
\label{timelike}	
t^{\mu} = \alpha u^{\mu} + \beta l^{\mu}
\end{align}
where the coefficients $\alpha$ and $\beta$ satisfy $\alpha^{2} \geq 2\alpha\beta u_{\mu}l^{\mu}$.  If we plug (\ref{timelike}) and (\ref{heatless fluid}) into (\ref{weak energy}) then we find
\begin{align}
\label{stress E time time}
T_{\mu\nu}t^{\mu}t^{\nu} = \rho_{m}\left(\alpha^{2} - 2\alpha\beta u_{\mu}l^{\mu}\right) + \beta^{2}\left[\left(\rho_{m} + p_{m}\right)\left(u_{\mu}l^{\mu}\right)^{2} + \pi_{\mu\nu}l^{\mu}l^{\nu}\right]\geq 0.
\end{align}
This implies
\begin{align}
\label{weak E no heat}
&\ \rho_{m} \geq 0\cr
&\left(\rho_{m} + p_{m}\right)\left(u_{\mu}l^{\mu}\right)^{2} + \pi_{\mu\nu}l^{\mu}l^{\nu} \geq 0.
\end{align}
The second condition is exactly what we got from the null energy conditions.  The weak energy condition, however, also requires the matter energy density be positive everywhere.

The last energy condition we consider is the strong energy condition, which imposes
\begin{align}
T_{\mu\nu}t^{\mu}t^{\nu} \geq \frac{1}{2} T^{\mu}{}_{\mu}t^{\nu}t_{\nu}.
\end{align}
It is easily shown this requires that the components (\ref{heatless fluid}) satisfy
\begin{align}
\frac{1}{2}\left(\rho_{m} + 3p_{m}\right)\left(\alpha^{2} - 2\alpha\beta u_{\mu}l^{\mu}\right) + \beta^{2}\left[\left(\rho_{m} + p_{m}\right)\left(u_{\mu}l^{\mu}\right)^{2} + \pi_{\mu\nu}l^{\mu}l^{\nu}\right]\geq 0
\end{align}
which implies
\begin{align}
\label{strong E no heat}
&\ \rho_{m} + 3p_{m} \geq 0\cr
&\left(\rho_{m} + p_{m}\right)\left(u_{\mu}l^{\mu}\right)^{2} + \pi_{\mu\nu}l^{\mu}l^{\nu} \geq 0.
\end{align}
The second inequality is again just the null energy condition.  We note the strong energy condition has the physical interpretation of demanding that gravity be attractive.

Our discussion so far applies to any fluid solution whose heat flux $q^{\mu} = 0$.  Let us now consider such solutions that also source spherically symmetric Kerr-Schild solutions.  We see from (\ref{Schwarzschild metric}) and (\ref{fluid comps}) that $\phi(r)$ and its derivatives parameterize these fluid solutions.  The energy conditions then become inequalities that constrain $\phi(r)$ and its derivatives.  To see this we first write the arbitrary null vector $l^{\mu}$ appearing in these conditions as 
\begin{align}
l^{\mu} = \left(l^{0}, \vec{l}\right).
\end{align}
For $l^{\mu}$ to be null with respect to (\ref{Schwarzschild metric}) its components must satisfy
\begin{align}
\label{null condition}
|\vec{l}| = \frac{1 - \phi}{\sqrt{1 - \phi~ {\rm sin}^{2}\alpha}}l^{0}
\end{align}
where ${\rm cos}\alpha = \hat{r}\cdot\hat{l}$.  Using (\ref{null condition}) and (\ref{fluid comps}) we can write the inequality (\ref{null E}) appearing in the null, weak and strong energy conditions as
\begin{align}
0 &\leq \left(\rho_{m} + p_m\right)\left(u_{\mu}l^{\mu}\right)^{2} + \pi_{\mu\nu}l^{\mu}l^{\nu}\cr
&=\left(\rho_{m} + p_m\right)\left[\left(1 - \phi\right)\left(l^{0}\right)^{2} - \frac{1}{2}\left(\frac{3 - \phi}{1 - \phi}{\rm cos}^{2}\alpha - 1\right)\left(\frac{\left(1 - \phi\right)^{2}}{1 - \phi ~{\rm sin}^{2}\alpha}\right)\left(l^{0}\right)^{2}\right]\cr
&= \left(\rho_{m} + p_m \right)\left(\frac{3}{2}\right)\left[{(l^0)^2 (1-\phi)^2{\rm sin}^2\alpha \over (1-\phi~ {\rm sin}^2 \alpha)}\right].
\end{align}
For $\phi(r) \textless 1$ the bracketed term on the last line is then greater than or equal to zero for all $r$.  To satisfy the null energy condition its components must then satisfy
\begin{align}
\label{null constraints}
\rho_{m} + p_m &\geq 0.
\end{align}
We see that the null energy condition allows for negative energy densities so long as the pressure is positive and compensates accordingly.  The weak energy condition then requires
\begin{equation}
\label{weak constraints}
\rho_{m} \geq 0 ~ {\rm and}~\rho_{m} + p_m \geq 0
\end{equation}
while the strong energy condition requires
\begin{equation}
\label{strong constraints}
\rho_{m} + 3p_m \geq 0~{\rm and}~\rho_{m} + p_m \geq 0.
\end{equation}
The unique solution to $\nabla^{2}\phi = -4\pi {\rho} = -\rho_{K}$ that satisfies (\ref{finite constraints}) is
\begin{align}
\label{solution}
\phi(r) = \frac{1}{r}\int_{0}^{r} dr' r'^{2}\rho_{K}(r') - \int_{0}^{r}dr'r'\rho_{K}(r')
\end{align}
which implies
\begin{align}
\label{matter density 1}
\rho_{m} = -\frac{1}{r^{2}}\int_{0}^{r}dr'r'\rho_{K}(r').
\end{align}
We can use (\ref{matter density 1}) to express the pressure in terms of $\rho_{K}$
\begin{align}
\label{pressures}
p_{m} = \frac{1}{3}\left(\rho_{K} + \frac{1}{r^{2}}\int_{0}^{r}dr'r'\rho_{K}(r')\right),
\end{align}
and write the null energy condition (\ref{null constraints}) as
\begin{align}
\label{null energy condition}
\rho_{K} - \frac{2}{r^{2}}\int_{0}^{r}dr'r'\rho_{K}(r') \geq 0.
\end{align}
Similarly we can write the weak energy condition (\ref{weak constraints}) as 
\begin{equation}
\label{weak energy condition}
-\frac{1}{r^{2}}\int_{0}^{r}dr'r'\rho_{K}(r')\geq 0~{\rm and}~\rho_{K} - \frac{2}{r^{2}}\int_{0}^{r}dr'r'\rho_{K}(r')\geq 0
\end{equation}
and the strong energy condition (\ref{strong constraints}) as
\begin{equation}
\label{strong energy condition}
\rho_{K} \geq 0~{\rm and}~\rho_{K} - \frac{2}{r^{2}}\int_{0}^{r}dr'r'\rho_{K}(r') \geq 0
\end{equation}
We immediately see that we cannot satisfy the strong and weak energy conditions, (\ref{strong energy condition}) and (\ref{weak energy condition}), at the same time for any $\rho_{K}(r)$.  Most sensible matter distributions are expected to satisfy both the strong and weak energy conditions.

Its worth mentioning that distributions such that $\rho_{K}(0) = 0$ cannot satisfy the weak energy condition.  To see this we go to the small $r$ limit and expand $\rho_{K}(r)$ as
\begin{align}
\rho_{K}(r) = \sum_{n}\frac{1}{n!}\rho_{K}^{(n)}(0) r^{n}
\end{align}
where $\rho_{K}^{(n)}(0)$ is the nth derivative of $\rho_{K}$ at the origin.  If we plug this expansion into (\ref{weak energy condition}) the weak energy condition becomes
\begin{align}
\label{sums and inequalities}
-\sum_{n}\frac{\rho_{K}^{(n)}(0)}{n!}\frac{1}{n+2}r^{n}&\geq 0\cr
\sum_{n}\frac{\rho_{K}^{(n)}(0)}{n!}\left[1 - \frac{2}{n+2}\right] r^{n} &\geq 0.
\end{align}
If $\rho_{K}(0) = 0$, then the leading order term becomes the linear one.  However, the first inequality in (\ref{sums and inequalities}) implies $\rho'_{K}(0) \leq 0$ while the second one implies $\rho'_{K}(0) \geq 0$.  These then force $\rho'_{K}(0) = 0$.  To satisfy (\ref{sums and inequalities}) order by order we must have $\rho_{K}^{(n)}(0) = 0$.

The weak energy condition excludes $\rho_{K}(r)$ that are positive for all $r$.  If $\rho_{K}(r)$ is to be positive at all and satisfy the weak energy condition then it must start off negative and flip sign as $r$ increases.  There are some $\rho_{K}(r)$ that satisfy the null energy condition (\ref{null energy condition}) and are postive for all $r$.  

\section{Concluding Remarks}

We have studied the nature of the sources ({\it i.e.} the stress tensors) for static Kerr-Schild metrics with spherical symmetry.  These metrics are double copies of spherically symmetric electrostatic potentials.  In most cases the stress-energy tensor is not a perfect fluid and contains shear stresses.  We also found that specifying the charge distribution sourcing the single copy and imposing $\nabla_{\mu}T^{\mu}_{~\nu}=0$ fixed these stress-energy tensors.  Finally we find that both the weak and strong energy conditions are never satisfied simultaneously by the stress-energy tensors arising from static spherically symmetric Kerr-Schild metrics with no curvature singularities or horizons.  Most sensible matter distributions are expected to satisfy these conditions.  It is possible, however, that there are other coordinate systems that display double copy behavior and map E\&M solutions to GR ones that satisfy the energy conditions \cite{Luna:2015paa}.

\acknowledgments

We thank Zvi Bern, Sean Carroll, Clifford Cheung and Donal O'Connell for useful discussions.  MBW would also like to thank Latham Boyle.  This work was supported by the Gordon and Betty Moore Foundation through Grant No.\
776 to the Caltech Moore Center for Theoretical Cosmology and Physics, and by
the DOE Grant DE-SC0011632. We are also grateful for the support provided by the Walter
Burke Institute for Theoretical Physics.

\end{document}